\documentclass[journal=aamick,manuscript=article]{achemso}

\setkeys{acs}{articletitle = true}
\usepackage{soul}
\usepackage{tabularx}
\usepackage[version=3]{mhchem} 
\usepackage[T1]{fontenc}       
\usepackage{epstopdf}
\DeclareGraphicsExtensions{.pdf,.eps,.png,.jpg,}
 
\usepackage{epstopdf}
\usepackage{multirow}
\usepackage{soul}
\usepackage[normalem]{ulem}
\usepackage{float}
\usepackage[utf8]{inputenc}
\usepackage[T1]{fontenc}
\usepackage{mathptmx}
\usepackage{etoolbox}
\usepackage{gensymb}
\usepackage{amsmath}
\usepackage{xcolor}
\usepackage{natbib}
\usepackage{alphabeta}
\usepackage{placeins}
\DeclareUnicodeCharacter{2009}{\,}
\usepackage[utf8]{inputenc}
\usepackage{hyperref,url}
\makeatletter
\def\@email#1#2{%
 \endgroup
 \patchcmd{\titleblock@produce}   {\frontmatter@RRAPformat}
  {\frontmatter@RRAPformat{\produce@RRAP{*#1\href{}{#2}}}\frontmatter@RRAPformat}
  {}{}
}%
\makeatother
  
\title{Monolithic  $\beta$-Ga$_2$O$_3$ Bidirectional MOSFET\\}
 \author{Pooja Sharma}
  \author{Saurabh Lodha}
 \email{slodha@ee.iitb.ac.in}
 \affiliation{ Department of Electrical Engineering, IIT Bombay, Mumbai, India - 400076
}%
\begin{document}\date{\today}

\begin{abstract}

We report a monolithic bidirectional dual-gate metal-oxide-semiconductor field effect transistor (MOSFET) fabricated on epitaxially grown $\beta$-Ga$_2$O$_3$, demonstrating efficient two-way conduction and blocking. It features two independently controlled gates and operates in four distinct modes, offering flexibility in managing current and voltage in the first and third quadrants. This versatility makes it ideal for various power conversion system applications. The device operates at a low negative threshold voltage (\textasciitilde{}-2.4 V for both gates) with a zero turn-on drain voltage and an on-resistance of approximately 500 $\Omega \cdot$mm. It exhibits a high on/off current ratio of $10^7$ in all three conducting modes. In the blocking mode, the device breakdown was measured to be more than $\pm$350 V with a current compliance of 0.5 mA/mm. The estimated breakdown field and power figure of merit for the device are 0.35 MV/cm and 1.6 MW/cm$^2$ respectively.
\end{abstract}

\section*{Introduction}
Bidirectional conductance is utilized in various applications such as AC-AC power conversion systems \cite{ac-ac_2002}, battery management systems in electric vehicles for efficient charging and discharging \cite{BiDi_EV2023, EV2012, hybridEV2010}, energy storage systems in smart grids \cite{Smartgrid2016} for effective energy supply and storage and renewable energy systems such as photovoltaic arrays \cite{PV2014, sustain_energy_2023}. It s also a key component of uninterrupted power supplies (UPS) \cite {UPS2010}, motor drives \cite{motordrives2019} in industrial machinery, and base stations for telecommunications \cite{ALAM2024100594}. 

Initially, thyristors with dedicated peripheral circuits were used to create bidirectional switches, resulting in inefficient and bulky power circuits \cite{bulky_thyristor_switch2002}. Over time, the use of power transistors has made bidirectional power switches (BPS) smaller and more suitable for complex power circuits\cite{Monolithic_Bidirectional_Power_Transistors}. Typically, BPSs consist of multiple discrete silicon-based insulated gate bipolar transistors (IGBTs) or power metal-oxide semiconductor field effect transistors (MOSFETs) arranged in common source, common drain, anti-parallel reverse blocking, or diode bridge configurations.\cite{Monolithic_Bidirectional_Power_Transistors} These commercially available bidirectional switches offer versatile functionality but come with limitations. These include complex control requirements \cite{complexcontrolBI2004}, large component count, design complexity and cost, high switching losses, and parasitic effects that can impact efficiency and performance. Despite these challenges, bidirectional switches are valuable for their flexibility and capability in managing bidirectional current flow.

The emergence of monolithic bidirectional power transistor switches (M-BPSs) marked a significant advancement\cite{emerging_Mono_BPS2023, Monolithic_switch_2023}. These transistors, fabricated with two gates, facilitated integration of control and power handling capabilities, improving performance characteristics such as faster switching speeds and lower power losses. Monolithic designs also reduced the footprint and complexity of power conversion modules \cite{Monolithic_switch_2023}, making them ideal for compact and high-density electronic applications. The realization of a monolithic bidirectional switch with dual-gate structure on GaN was reported by Morita et al. \cite{BiGaN_IEDM}, featuring a normally-off design. Subsequently, Peng et al. \cite{peng2024monolithic} demonstrated a monolithic fully-controlled high electron mobility transistor (HEMT) BPS. This design further reduced power loss, featuring a low turn-on voltage and high threshold voltage ($V_{TH}$) through the use of embedded Schottky diodes. The dual-gate bidirectional switch takes advantage of a common drain region to keep the device’s cell pitch as small as possible.\cite{Monolithic_switch_2023} Panasonic offers a normally-off dual-gate monolithic GaN bidirectional switch that delivers symmetric 100 A conduction and 1,100 V blocking voltage \cite{Panasonic2023}. 

Recent developments in SiC-based M-BPSs have resulted in higher blocking voltages compared to GaN-based devices, which typically achieve up to $\pm$650 V. For example, SiC vertical bidirectional FETs can achieve $\pm$1200 V but involve a common-drain setup with large chip area.\cite{han2020monolithic} A true monolithic 4H-SiC IGBT, with a shared drain region has shown blocking voltages up to $\pm$7 kV. However, these vertical SiC M-BPS architectures require complex double-sided lithography and present cooling challenges \cite{chowdhury2016operating}. The more favourable dual lateral gate M-BPS design has not been explored on SiC. On the other hand, most commercially available lateral GaN transistors target the 650 V rating \cite{chen2017gan, transform_GaN} and a kV-range GaN transistor is not available commercially \cite{GaN_status_2021}. However, single-gate  $\beta$-Ga$_2$O$_3$ transistors are more amenable to lateral designs and have shown better high voltage (kV-range) handling capabilities compared to both lateral GaN and lateral SiC devices\cite{Bhattacharyya_2022}. Therefore, exploring bidirectional transistor capabilities with $\beta$-Ga$_2$O$_3$ is worthwhile.





We demonstrate a simple monolithic bidirectional MOSFET designed and fabricated on epitaxially grown $\beta$-Ga$_2$O$_3$. Both gates in the transistor exhibit nearly the same $V_{TH}$ (-2.4 V and -2.3 V), with an on/off-current ($I_{ON}$/$I_{OFF}$) ratio of approximately $10^7$ at a drain bias ($V_{DS}$) of 15 V (in saturation region) for both gates. The dual-gate MOSFET is tested in four distinct modes of operation by applying different gate biasing schemes. The four modes include bidirectional conduction, bidirectional blocking, and two diode modes. These tests show its ability to manage different currents and blocking voltages as needed. In the two diode modes, the device has non-zero (\textasciitilde{}$\pm$0.8 V) turn-on voltages due to the formation of a diode junction in the channel because of two different bias voltages applied to the two gates. But in the bidirectional conduction mode, the transistor turns-on at zero volts, which would help in cutting down the power loss. On-resistance ($R_{ON}$) of \textasciitilde{}500 $\ohm$.mm is obtained for the two transistors in series during the bidirectional conduction mode. In the blocking mode, it shows voltage blocking capability of more than $\pm$350 V. These parameters highlight its suitability for compact and efficient power conversion applications. 

\section*{Experimental}
\begin{figure}
\includegraphics[width=1\linewidth]{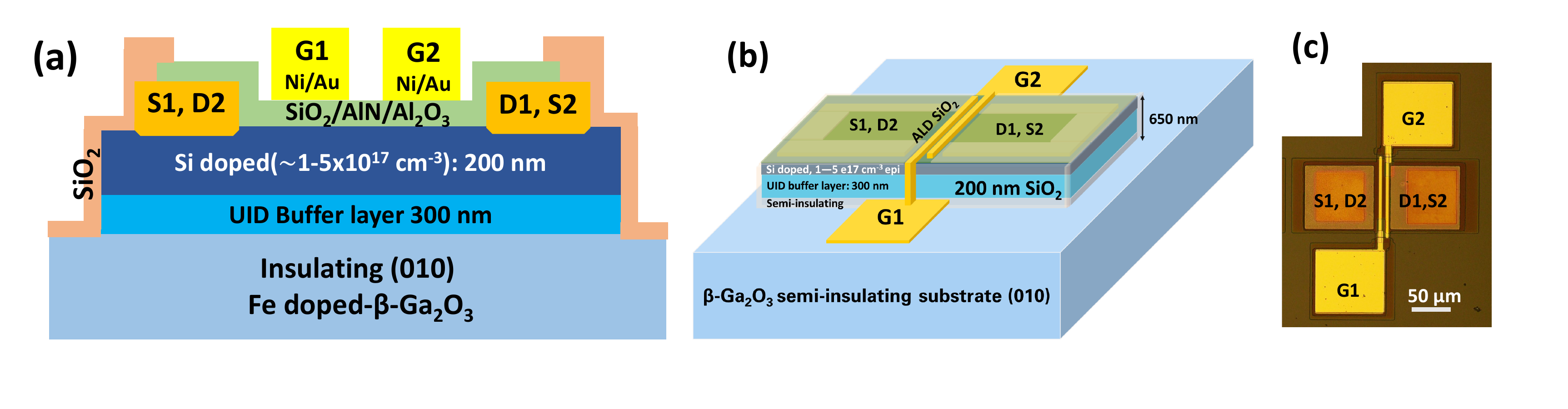}
\caption{\label{fig:fig1}(a) Cross-section, (b) microscope image and (c) 3D schematic of the monolithic bidirectional MOSFET with $L_G$= 2 $\mu$m, $L_{S1-S2}$= 15 $\mu$m, $L_{GD}$= 10 $\mu$m, and a common drain region of 5 $\mu$m.}

\end{figure}

The 5 x 5 mm, epi-grown, and $(010)$ oriented $\beta$-Ga$_2$O$_3$ substrate used in this study was procured from Novel Crystal Technology Inc. Molecular beam epitaxy grown stack used for the device consists of 200 nm thick, $1-5$ x $10^{17}$ cm$^{-3}$ Si-doped and 300 nm thick unintentionally doped (UID) layers on top of an Fe-doped semi-insulating substrate as shown in Fig. \ref{fig:fig1}. 
It was cleaned using a standard cleaning procedure \cite{Biswas2019, psharma2024,PSedtm2024}. This involved ultrasonication in methanol and acetone for 3 minutes to degrease, followed by immersion in piranha solution for 5 minutes to eliminate organic residues. Next, we used an optical lithography system, microwriter from Durham Magneto Optics Ltd., to define the source-drain (S-D) regions. 
$\beta$-Ga$_2$O$_3$ was then etched in the S-D regions to create \textasciitilde{}10 nm recess using BCl$_3$/Cl$_2$/Ar inductively coupled plasma reactive ion etching (ICPRIE) at a chamber pressure of 2 Pa, ICP power of 400 W and RF power of 75 W for 100 s. RIE treatment has been reported to significantly reduce the S-D contact resistance\cite{Masataka_2012_RIE_etch}. This was done prior to the evaporation and lift-off of a Ti:Au (30/100 nm) metal stack. To further reduce the contact resistance, the Ti/Au stack was annealed at 470 °C for 3 minutes in an N$_2$ atmosphere using a rapid thermal annealing (RTA) system. 

\FloatBarrier
\FloatBarrier

The gate dielectric stack was deposited using an atomic layer deposition (ALD) system. It consists of SiO$_2$ (5 nm), AlN (5 nm), and Al$_2$O$_3$ (12 nm) \cite{biswas2020} layers deposited in a sequential \textit{in-situ} fashion. SiO$_2$ was deposited using O$_2$ plasma at 250 \degree C, AlN with NH$_3$ plasma at 200 \degree C, and Al$_2$O$_3$ through thermal ALD at 250 \degree C.  
MESA isolation of the active transistor region was obtained by etching using an Ar-ion milling system, achieving an etch depth of approximately 650 nm, exposing/reaching the semi-insulating substrate.
The source-drain and MESA sidewalls were then coated with 200 nm of sputtered SiO$_2$ using optical lithography and lift-off. Finally, gate metal Ni/Au (30/100 nm) was deposited using an e-beam evaporator. The fabricated devices were electrically characterized using the B1500 and B1505 semiconductor device analyzers from Keysight. \\

\section*{Results and Discussion}
Conventional unidirectional as well as  bidirectional MOSFETs were fabricated together in the same process run on the same substrate to ensure a fair and direct comparison of their electrical performance, with identical gate stack and source-drain processing. The single-gate, unidirectional MOSFET helped in verifying unit transistor characteristics using the triple dielectric layered gate stack.

\begin{figure}[ht]
\includegraphics[width=1\linewidth]{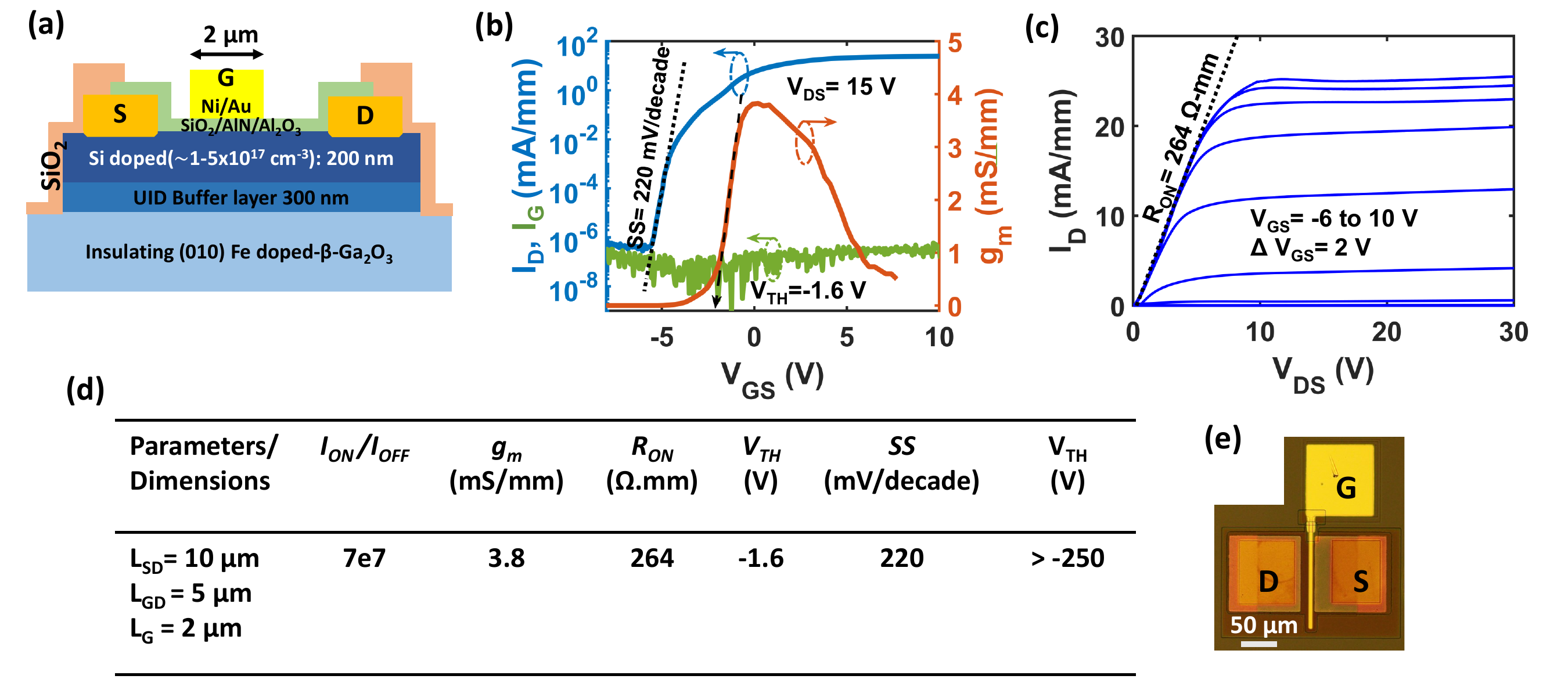}
\caption{\label{fig:fig2}(a) Cross-section with $L_G$= 2 $\mu$m, $L_{SD}$= 10 $\mu$m, $L_{GD}$= 5 $\mu$m, (b) transfer characteristics, transconductance (g$_m$) and (c) output characteristics of fabricated conventional single-gate unidirectional $\beta$-Ga$_2$O$_3$ MOSFET,  
(d) table containing extracted parameters, and (e) microscope image of fabricated device.}
\end{figure}

Transfer (forward and reverse sweeps are shown in supplementary material, section 1) and output characteristics of  a single-gate MOSFET ($L_G$= 2 $\mu$m, $L_{SD}$= 10 $\mu$m, $L_{GD}$= 5 $\mu$m, Fig.  \ref{fig:fig2}(a)) are shown in Fig.  \ref{fig:fig2}(b) and (c) respectively. The characteristics were measured at room temperature.
The device turns-on at a $V_{TH}$ of -1.6 V, indicating depletion mode operation. MESFETs on similar substrates (in terms of doping, epi-layer thickness and orientation) have been reported with $V_{TH}$ values of nearly \textasciitilde{}-20 V \cite{MKuball2018_Vth, MHWong2016_Vth, MHWong2013_Vth_BV}. The dielectric stack was chosen to increase the $V_{TH}$ and move the device operation closer to normally-off state. The increase can be attributed to a positive flat band voltage for MOS capacitors fabricated with the SiO$_2$/AlN/Al$_2$O$_3$ dielectric stack reported by Biswas et al. \cite{biswas2020}. SiO$_2$ prevents gate leakage due to its high conduction band offset ($\Delta$$E_C$ \textasciitilde{}3.1 eV) \cite{konishi2016large} and forms a good interface with the lowest reported interface trap density, $D_{IT}$ ($5.1$ x $10^{10}$ cm$^{-2}$eV$^{-1}$)  on $\beta$-Ga$_2$O$_3$ \cite{Biswas2019, zeng2016interface}. Al$_2$O$_3$ acts as the blocking layer, while AlN serves as the charge trapping layer, with a smaller conduction band offset ($\Delta$$E_C$ \textasciitilde{}1.4 eV) with $\beta$-Ga$_2$O$_3$. The stack altogether shifts $V_{TH}$ towards 0 V.  Although the stack can be used to obtain enhancement mode operation through charge trapping, we have not tested charge trapping in this study.
Extracted subthreshold slope (SS) and transconductance (g$_ m$) values are 220 mV/decade and 3.82 mS/mm respectively. The $I_{ON}$/$I_{OFF}$ obtained is \textasciitilde{}7 x 10$^7$ at $V_{DS}$= 15 V with a maximum on-current of 24 mA/mm (at gate bias, $V_{GS}$, of 10 V) and off-current of nearly 1 nA/mm. The single-gate transistor exhibits a breakdown voltage of more than 250 V (supplementary material, section 2), consistent with reported values on similar substrates,\cite{Uttam2018_BV, MHWong2013_Vth_BV} without incorporating any field management techniques. Extracted key parameter values have been tabulated in Fig. \ref{fig:fig2}(d). Contact and channel resistance values extracted from transfer length measurements (TLM) are 5.14 k$\Omega$.mm and 180 k$\Omega$/sq (supplementary material, section 3) respectively. The $R_{ON}$ value from output characteristics is \textasciitilde{}264.3  $\Omega$.mm. The higher contact resistance accounts for the slightly lower current values compared to existing transistor reports on similarly doped substrates.\cite{MHWong2013_Vth_BV, MHWong2016_Vth}

\FloatBarrier
\FloatBarrier
\begin{figure}[h]
\includegraphics[width=1\linewidth]{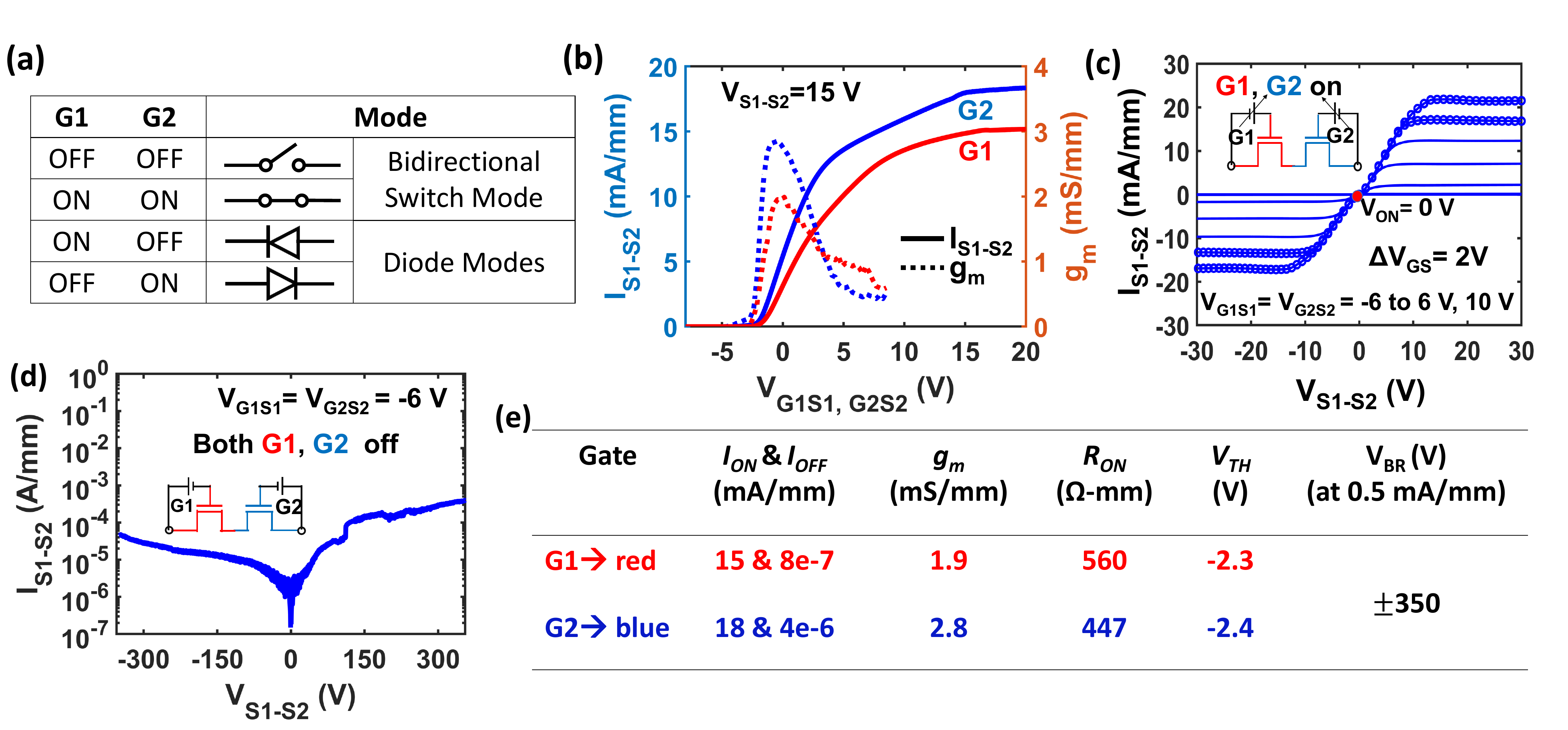}
\caption{\label{fig:fig3}(a) Four distinct operation modes of the monolithic bidirectional MOSFET, (b) individual transfer characteristics of the two gates of the bidirectional transistor, (c) output characteristics with both gates at the same voltage in on-condition, (d) off-condition with both gates at -6 V, and (e) parameter values extracted from transfer and output characteristics.}
\end{figure}
\FloatBarrier
The monolithic dual-gate MOSFET structure shown in Fig. \ref{fig:fig1} allows for a bidirectional switch without significant active area penalty, as the drift layer is shared by two transistors.  
This device structure was fabricated and characterized by biasing it in four different modes. Two of the modes are switch modes (bidirectional blocking and bidirectional conduction) and other two are diode modes as illustrated in Fig. \ref{fig:fig3}(a). 
For the first gate (G1), the source and drain electrodes are designated as S1 and D1. Similarly, for the second gate (G2), the source and drain electrodes are S2 (same as D1) and D2 (same as S1) respectively.

Transfer characteristics of the bidirectional MOSFET are shown in Fig. \ref{fig:fig3}(b) where both gates exhibit nearly the same $V_{TH}$ of \textasciitilde{}-2.4 V. These characteristics were measured with G1-to- S1 voltage (V$_{G1-S1}$) swept from -8 to 20 V with G2-to-S2 voltage (V$_{G2-S1}$) at 10 V and vice versa. The 0.1 V difference in threshold voltages ($V_{TH-G1}$ and $V_{TH-G2}$) for the two unit transistors is possibly due to slight misalignment or asymmetric placement of gates between the source and the drain. $g_m$ and $I_{ON}$ values for the two gates also show a small disparity ($\Delta g_m$ \textasciitilde{}0.9 mS/mm and $\Delta$$I_{ON}$ \textasciitilde{}3 mA/mm), again due to asymmetric placement of the two gates.

\FloatBarrier
\begin{figure}
\includegraphics[width=0.75\linewidth]{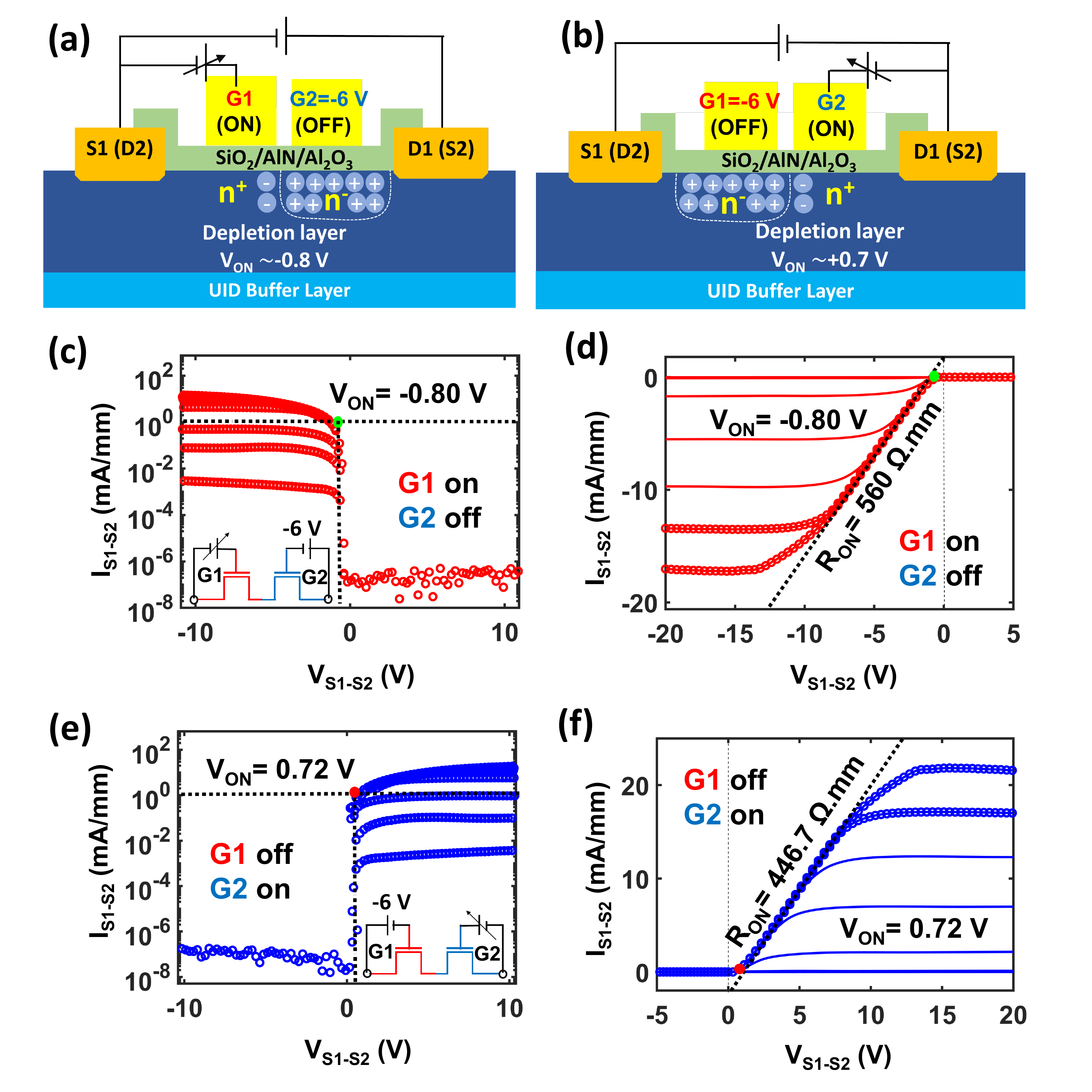}
\caption{\label{fig:fig4}
Electrostatics of the bidirectional MOSFET is demonstrated with the device biased to operate in two diode modes where (a) G1 is on and G2 is off, and (b) G1 is off and G2 is on, illustrating how the channel junction influences the turn-on voltage in the two diode modes. The output characteristics of the bidirectional MOSFET in reverse diode mode are shown on (c) logarithmic scale and (d) linear scale, and in forward diode mode on  (e) logarithmic scale and (f) linear scale, with one gate in the on condition and the other at -6 V (off-state). These measurements are for a fabricated bidirectional MOSFET with gate length ($L_G$) of 2 $\mu$m, source-drain length ($L_{SD}$) of 15 $\mu$m, gate-drain length ($L_{GD}$) of 10 $\mu$m, and a common drain region of 5 $\mu$m. }

\end{figure}
\FloatBarrier

Output characteristics of the bidirectional transistor, shown in Fig. \ref{fig:fig3}(c), demonstrate 0 V turn-on operation and bidirectional current conduction with both gates (G1 and G2) biased in the on state,  $V_{G1-S1}$ > $V_{TH-G1}$ and $V_{G2-S2}$ > $V_{TH-G2}$.  This can significantly reduce conduction losses. In bidirectional blocking (off) mode, both gates are biased at -6 V i.e. $V_{G1-S1}$ < $V_{TH-G1}$ and $V_{G2-S2}$ < $V_{TH-G2}$, and S-D bias is swept from -350 V to 350 V as shown in Fig. \ref{fig:fig3}(d). The current levels remain below $10^{-3}$ A (in the $\mu$A/mm range) upto $\pm$350 V. 
Key parameters extracted from the transfer and output characteristics are summarized in Fig. \ref{fig:fig3}(e).


%

In diode mode operation, the transistor can be biased to work in reverse or forward blocking modes with charge distribution in channel as shown in Fig. \ref{fig:fig4}(a) and (b) respectively. In the reverse conduction mode, where G1 is on and G2 is off, current $I_{S1-S2}$ flows from terminal S2 (or D1) to terminal D2 (or S1).  When $V_{G1-S1}$ > $V_{TH-G1}$ and $V_{G2-S2}$ < $V_{TH-G2}$, the $\beta$-Ga$_2$O$_3$ region under G1 is in accumulation while that under G2 is in depletion as illustrated in Fig. \ref{fig:fig4}(a). This creates a lateral junction in the $\beta$-Ga$_2$O$_3$ channel region, similar to a p-n junction, contributing to the turn-on voltage as depicted by $I-V$ characteristics shown in Fig. \ref{fig:fig4}(c) and (d) for forward conduction mode. 
Similarly, a voltage drop occurs in forward conduction mode as well, with reversed polarity on the two gates. In this mode, the current $I_{S2-S1}$ flows from terminal S1 (D2) to terminal D1 (S2) when $V_{G1-S1}$ < $V_{TH-G1}$ and $V_{G2-S2}$ > $V_{TH-G2}$ as illustrated in Fig. \ref{fig:fig4}(e) and (f). Both diode modes show an on/off-current ratio of nearly $10^7$. The turn-on voltages are -0.8 V and +0.7 V for reverse and forward diode-like modes, respectively, with corresponding $R_{ON}$ of 560 $\Omega$·mm and 447 $\Omega$·mm. The $R_{ON}$ value is nearly twice that of the single-gate transistor, possibly due to two transistors in series with extended channel length ($\Delta$$L_{SD}$ \textasciitilde{}5 $\mu$m).  Extracted breakdown field ($E_{BR}$) values are 0.35 MV/cm for the bidirectional transistor ($V_{BR}$ > 350 V and $L_{GD}$= 10 $\mu$m) and 0.5 MV/cm  for the single-gate unidirectional transistor (with $V_{BR}$ > 250 V and $L_{GD}$= 5 $\mu$m). The power figure of merit (PFOM) values for the two devices, bidirectional MOSFET ($R_{on,sp}$ \textasciitilde{}75 m$\Omega$.cm$^2$) and unidirectional single-gate ($R_{on,sp}$ \textasciitilde{}26 m$\Omega$.cm$^2$), are calculated to be approximately 1.6 and 2.4 MW/cm$^2$ respectively. Here, the specific-on-resistance, $R_{on,sp}$ is obtained by normalising the $R_{ON}$ with channel length of the device, $L_{SD}$.
The non-zero turn-on voltage can contribute to conduction losses when the device operates in either of these two diode modes. This bidirectional switch has no on-state voltage offsets, enabling low-loss operation, but the slightly negative $V_{TH}$ i.e. depletion mode operation can contribute to power loss. 



\section*{Conclusion}
We have demonstrated a monolithic dual-gate bidirectional MOSFET using $\beta$-Ga$_2$O$_3$. The transistor operates in depletion mode with a small negative threshold voltage of -2.4 V and a high $I_{ON}$/$I_{OFF}$ ratio of $10^7$ at $V_{DS}$=15 V for both gates with breakdown voltages of $\pm$350 V. The four modes of operation—bidirectional conduction, bidirectional blocking, and two diode modes—highlight the versatility of the transistor. Zero turn-on voltage in the bidirectional conduction mode could significantly reduce power loss, enhancing overall efficiency. However, the depletion mode operation can inherently contribute to conduction losses. Future work can address this limitation by fabricating bidirectional MOSFETs with enhancement mode operation, further optimizing performance. The blocking mode performance can also be enhanced with the use of field plates to increase the breakdown voltages. Additionally, on-current can be improved by employing techniques such as regrown source-drain contacts during the fabrication process, which help reduce contact resistance.

In conclusion, this bidirectional transistor with its four modes of operation in a lateral compact design furthers the potential of  $\beta$-Ga$_2$O$_3$ for kV-class bidirectional power switches in a host of power electronic applications such as motor drive systems, hybrid vehicles, battery storage, super capacitor charging, UPS systems and solar power generation. 

\section*{Supplementary Material}
The supplementary material comprises TLM measurement data and double-sweep (forward and reverse) hysteresis data for single-gate, unidirectional MOSFET. 

\section*{Acknowledgement}

The authors thank Dr. Poulomi Chakrabarty, Netaji Suvachintak and Kanchan Singh Rana for help with Ar-ion milling, ICPRIE etch and e-beam evaporation for contact deposition respectively. The authors acknowledge the Indian Institute of Technology Bombay Nano-fabrication Facility (IITBNF) for the usage of its device fabrication and characterization facilities and support from the Ministry of Electronics and Information Technology (through project 5(1)/2017-NANO) and Department of Science and Technology (through project DST/NM/NNetRA/2018(G)-IIT-B), Government of India, for funding this work. Pooja Sharma thanks University Grants Commission (UGC) for the doctoral fellowship. 

\section*{Data Availability}
The data supporting this study's findings are included within the article [and its supplementary material]. \\
\bibliography{manuscript}

\providecommand{\noopsort}[1]{}\providecommand{\singleletter}[1]{#1}%
\providecommand{\latin}[1]{#1}
\makeatletter
\providecommand{\doi}
  {\begingroup\let\do\@makeother\dospecials
  \catcode`\{=1 \catcode`\}=2 \doi@aux}
\providecommand{\doi@aux}[1]{\endgroup\texttt{#1}}
\makeatother
\providecommand*\mcitethebibliography{\thebibliography}
\csname @ifundefined\endcsname{endmcitethebibliography}  {\let\endmcitethebibliography\endthebibliography}{}
\begin{mcitethebibliography}{36}
\providecommand*\natexlab[1]{#1}
\providecommand*\mciteSetBstSublistMode[1]{}
\providecommand*\mciteSetBstMaxWidthForm[2]{}
\providecommand*\mciteBstWouldAddEndPuncttrue
  {\def\EndOfBibitem{\unskip.}}
\providecommand*\mciteBstWouldAddEndPunctfalse
  {\let\EndOfBibitem\relax}
\providecommand*\mciteSetBstMidEndSepPunct[3]{}
\providecommand*\mciteSetBstSublistLabelBeginEnd[3]{}
\providecommand*\EndOfBibitem{}
\mciteSetBstSublistMode{f}
\mciteSetBstMaxWidthForm{subitem}{(\alph{mcitesubitemcount})}
\mciteSetBstSublistLabelBeginEnd
  {\mcitemaxwidthsubitemform\space}
  {\relax}
  {\relax}

\bibitem[Wheeler \latin{et~al.}(2002)Wheeler, Rodriguez, Clare, Empringham, and Weinstein]{ac-ac_2002}
Wheeler,~P.; Rodriguez,~J.; Clare,~J.; Empringham,~L.; Weinstein,~A. Matrix converters: a technology review. \emph{IEEE Transactions on Industrial Electronics} \textbf{2002}, \emph{49}, 276--288\relax
\mciteBstWouldAddEndPuncttrue
\mciteSetBstMidEndSepPunct{\mcitedefaultmidpunct}
{\mcitedefaultendpunct}{\mcitedefaultseppunct}\relax
\EndOfBibitem
\bibitem[Wei and Li(2023)Wei, and Li]{BiDi_EV2023}
Wei,~C.; Li,~X. Review of Bidirectional DC-DC Converters for Electric Vehicle Energy Management Systems. 2023 7th International Conference on Smart Grid and Smart Cities (ICSGSC). 2023; pp 432--436\relax
\mciteBstWouldAddEndPuncttrue
\mciteSetBstMidEndSepPunct{\mcitedefaultmidpunct}
{\mcitedefaultendpunct}{\mcitedefaultseppunct}\relax
\EndOfBibitem
\bibitem[Onar \latin{et~al.}(2012)Onar, Kobayashi, Erb, and Khaligh]{EV2012}
Onar,~O.~C.; Kobayashi,~J.; Erb,~D.~C.; Khaligh,~A. A bidirectional high-power-quality grid interface with a novel bidirectional noninverted buck--boost converter for PHEVs. \emph{IEEE transactions on vehicular technology} \textbf{2012}, \emph{61}, 2018--2032\relax
\mciteBstWouldAddEndPuncttrue
\mciteSetBstMidEndSepPunct{\mcitedefaultmidpunct}
{\mcitedefaultendpunct}{\mcitedefaultseppunct}\relax
\EndOfBibitem
\bibitem[Du \latin{et~al.}(2010)Du, Zhou, Bai, Lukic, and Huang]{hybridEV2010}
Du,~Y.; Zhou,~X.; Bai,~S.; Lukic,~S.; Huang,~A. Review of non-isolated bi-directional DC-DC converters for plug-in hybrid electric vehicle charge station application at municipal parking decks. 2010 Twenty-Fifth Annual IEEE Applied Power Electronics Conference and Exposition (APEC). 2010; pp 1145--1151\relax
\mciteBstWouldAddEndPuncttrue
\mciteSetBstMidEndSepPunct{\mcitedefaultmidpunct}
{\mcitedefaultendpunct}{\mcitedefaultseppunct}\relax
\EndOfBibitem
\bibitem[Zhang and Chau(2016)Zhang, and Chau]{Smartgrid2016}
Zhang,~Z.; Chau,~K.-T. Pulse-width-modulation-based electromagnetic interference mitigation of bidirectional grid-connected converters for electric vehicles. \emph{IEEE Transactions on Smart Grid} \textbf{2016}, \emph{8}, 2803--2812\relax
\mciteBstWouldAddEndPuncttrue
\mciteSetBstMidEndSepPunct{\mcitedefaultmidpunct}
{\mcitedefaultendpunct}{\mcitedefaultseppunct}\relax
\EndOfBibitem
\bibitem[Chao and Huang(2014)Chao, and Huang]{PV2014}
Chao,~K.-H.; Huang,~C.-H. Bidirectional DC--DC soft-switching converter for stand-alone photovoltaic power generation systems. \emph{IET Power Electronics} \textbf{2014}, \emph{7}, 1557--1565\relax
\mciteBstWouldAddEndPuncttrue
\mciteSetBstMidEndSepPunct{\mcitedefaultmidpunct}
{\mcitedefaultendpunct}{\mcitedefaultseppunct}\relax
\EndOfBibitem
\bibitem[Sutikno \latin{et~al.}(2023)Sutikno, Aprilianto, and Purnama]{sustain_energy_2023}
Sutikno,~T.; Aprilianto,~R.~A.; Purnama,~H.~S. {Application of non-isolated bidirectional DC–DC converters for renewable and sustainable energy systems: a review}. \emph{Clean Energy} \textbf{2023}, \emph{7}, 293--311\relax
\mciteBstWouldAddEndPuncttrue
\mciteSetBstMidEndSepPunct{\mcitedefaultmidpunct}
{\mcitedefaultendpunct}{\mcitedefaultseppunct}\relax
\EndOfBibitem
\bibitem[Naden and Bax(2010)Naden, and Bax]{UPS2010}
Naden,~M.; Bax,~R. Generator with DC boost and split bus bidirectional DC-to-DC converter for uninterruptible power supply system or for enhanced load pickup. 2010; US Patent 7,786,616\relax
\mciteBstWouldAddEndPuncttrue
\mciteSetBstMidEndSepPunct{\mcitedefaultmidpunct}
{\mcitedefaultendpunct}{\mcitedefaultseppunct}\relax
\EndOfBibitem
\bibitem[Gorji \latin{et~al.}(2019)Gorji, Sahebi, Ektesabi, and Rad]{motordrives2019}
Gorji,~S.~A.; Sahebi,~H.~G.; Ektesabi,~M.; Rad,~A.~B. Topologies and Control Schemes of Bidirectional DC–DC Power Converters: An Overview. \emph{IEEE Access} \textbf{2019}, \emph{7}, 117997--118019\relax
\mciteBstWouldAddEndPuncttrue
\mciteSetBstMidEndSepPunct{\mcitedefaultmidpunct}
{\mcitedefaultendpunct}{\mcitedefaultseppunct}\relax
\EndOfBibitem
\bibitem[Alam \latin{et~al.}(2024)Alam, Minai, and Bakhsh]{ALAM2024100594}
Alam,~M.~A.; Minai,~A.~F.; Bakhsh,~F.~I. Isolated bidirectional DC-DC Converter: A topological review. \emph{e-Prime - Advances in Electrical Engineering, Electronics and Energy} \textbf{2024}, \emph{8}, 100594\relax
\mciteBstWouldAddEndPuncttrue
\mciteSetBstMidEndSepPunct{\mcitedefaultmidpunct}
{\mcitedefaultendpunct}{\mcitedefaultseppunct}\relax
\EndOfBibitem
\bibitem[Wheeler \latin{et~al.}(2002)Wheeler, Rodriguez, Clare, Empringham, and Weinstein]{bulky_thyristor_switch2002}
Wheeler,~P.; Rodriguez,~J.; Clare,~J.; Empringham,~L.; Weinstein,~A. Matrix converters: a technology review. \emph{IEEE Transactions on Industrial Electronics} \textbf{2002}, \emph{49}, 276--288\relax
\mciteBstWouldAddEndPuncttrue
\mciteSetBstMidEndSepPunct{\mcitedefaultmidpunct}
{\mcitedefaultendpunct}{\mcitedefaultseppunct}\relax
\EndOfBibitem
\bibitem[Huber and Kolar(2023)Huber, and Kolar]{Monolithic_Bidirectional_Power_Transistors}
Huber,~J.; Kolar,~J.~W. Monolithic Bidirectional Power Transistors. \emph{IEEE Power Electronics Magazine} \textbf{2023}, \emph{10}, 28--38\relax
\mciteBstWouldAddEndPuncttrue
\mciteSetBstMidEndSepPunct{\mcitedefaultmidpunct}
{\mcitedefaultendpunct}{\mcitedefaultseppunct}\relax
\EndOfBibitem
\bibitem[Itoh \latin{et~al.}(2004)Itoh, Sato, Odaka, Ohguchi, Kodatchi, and Eguchi]{complexcontrolBI2004}
Itoh,~J.; Sato,~I.; Odaka,~A.; Ohguchi,~H.; Kodatchi,~H.; Eguchi,~N. A novel approach to practical matrix converter motor drive system with reverse blocking IGBT. 2004 IEEE 35th Annual Power Electronics Specialists Conference (IEEE Cat. No.04CH37551). 2004; pp 2380--2385 Vol.3\relax
\mciteBstWouldAddEndPuncttrue
\mciteSetBstMidEndSepPunct{\mcitedefaultmidpunct}
{\mcitedefaultendpunct}{\mcitedefaultseppunct}\relax
\EndOfBibitem
\bibitem[Bindra(2023)]{emerging_Mono_BPS2023}
Bindra,~A. Emerging Monolithic Bidirectional Switches Bring New Energy to WBG Devices [From the Editor]. \emph{IEEE Power Electronics Magazine} \textbf{2023}, \emph{10}, 4--8\relax
\mciteBstWouldAddEndPuncttrue
\mciteSetBstMidEndSepPunct{\mcitedefaultmidpunct}
{\mcitedefaultendpunct}{\mcitedefaultseppunct}\relax
\EndOfBibitem
\bibitem[Veliadis(2023)]{Monolithic_switch_2023}
Veliadis,~V. Monolithic Bidirectional WBG Switches Rekindle Power Electronics Technology [Expert View]. \emph{IEEE Power Electronics Magazine} \textbf{2023}, \emph{10}, 71--75\relax
\mciteBstWouldAddEndPuncttrue
\mciteSetBstMidEndSepPunct{\mcitedefaultmidpunct}
{\mcitedefaultendpunct}{\mcitedefaultseppunct}\relax
\EndOfBibitem
\bibitem[Morita \latin{et~al.}(2007)Morita, Yanagihara, Ishida, Hikita, Kaibara, Matsuo, Uemoto, Ueda, Tanaka, and Ueda]{BiGaN_IEDM}
Morita,~T.; Yanagihara,~M.; Ishida,~H.; Hikita,~M.; Kaibara,~K.; Matsuo,~H.; Uemoto,~Y.; Ueda,~T.; Tanaka,~T.; Ueda,~D. 650 V 3.1 mΩ.cm$^2$ GaN-based monolithic bidirectional switch using normally-off gate injection transistor. 2007 IEEE International Electron Devices Meeting. 2007; pp 865--868\relax
\mciteBstWouldAddEndPuncttrue
\mciteSetBstMidEndSepPunct{\mcitedefaultmidpunct}
{\mcitedefaultendpunct}{\mcitedefaultseppunct}\relax
\EndOfBibitem
\bibitem[Peng \latin{et~al.}(2024)Peng, Mao, Xu, Yang, Peng, Yu, Zhang, Zhang, Zhao, Zhang, \latin{et~al.} others]{peng2024monolithic}
Peng,~G.; Mao,~W.; Xu,~S.; Yang,~C.; Peng,~Y.; Yu,~L.; Zhang,~Y.; Zhang,~T.; Zhao,~S.; Zhang,~C.; others Monolithic fully-controlled HEMT bidirectional power switch with merged Schottky barrier diodes and p-GaN gate transistors for ultra-low turn-on voltage and high threshold voltage. \emph{Japanese Journal of Applied Physics} \textbf{2024}, \emph{63}, 03SP11\relax
\mciteBstWouldAddEndPuncttrue
\mciteSetBstMidEndSepPunct{\mcitedefaultmidpunct}
{\mcitedefaultendpunct}{\mcitedefaultseppunct}\relax
\EndOfBibitem
\bibitem[Pan()]{Panasonic2023}
SiC and GaN usage in bidirectional power applications. \url{https://www.powerelectronicsnews.com/sic-and-gan-usage-in-bidirectional-power-applications/}\relax
\mciteBstWouldAddEndPuncttrue
\mciteSetBstMidEndSepPunct{\mcitedefaultmidpunct}
{\mcitedefaultendpunct}{\mcitedefaultseppunct}\relax
\EndOfBibitem
\bibitem[Han \latin{et~al.}(2020)Han, Agarwal, Kanale, Baliga, Bhattacharya, Cheng, Hopkins, Amarasinghe, and Ransom]{han2020monolithic}
Han,~K.; Agarwal,~A.; Kanale,~A.; Baliga,~B.~J.; Bhattacharya,~S.; Cheng,~T.-H.; Hopkins,~D.; Amarasinghe,~V.; Ransom,~J. Monolithic 4-terminal 1.2 kV/20 A 4H-SiC bi-directional field effect transistor (BiDFET) with integrated JBS diodes. 2020 32nd International Symposium on Power Semiconductor Devices and ICs (ISPSD). 2020; pp 242--245\relax
\mciteBstWouldAddEndPuncttrue
\mciteSetBstMidEndSepPunct{\mcitedefaultmidpunct}
{\mcitedefaultendpunct}{\mcitedefaultseppunct}\relax
\EndOfBibitem
\bibitem[Chowdhury \latin{et~al.}(2016)Chowdhury, Hitchcock, Stum, Dahal, Bhat, and Chow]{chowdhury2016operating}
Chowdhury,~S.; Hitchcock,~C.~W.; Stum,~Z.; Dahal,~R.~P.; Bhat,~I.~B.; Chow,~T.~P. Operating principles, design considerations, and experimental characteristics of high-voltage 4H-SiC bidirectional IGBTs. \emph{IEEE Transactions on Electron Devices} \textbf{2016}, \emph{64}, 888--896\relax
\mciteBstWouldAddEndPuncttrue
\mciteSetBstMidEndSepPunct{\mcitedefaultmidpunct}
{\mcitedefaultendpunct}{\mcitedefaultseppunct}\relax
\EndOfBibitem
\bibitem[Chen \latin{et~al.}(2017)Chen, H{\"a}berlen, Lidow, lin Tsai, Ueda, Uemoto, and Wu]{chen2017gan}
Chen,~K.~J.; H{\"a}berlen,~O.; Lidow,~A.; lin Tsai,~C.; Ueda,~T.; Uemoto,~Y.; Wu,~Y. GaN-on-Si power technology: Devices and applications. \emph{IEEE Transactions on Electron Devices} \textbf{2017}, \emph{64}, 779--795\relax
\mciteBstWouldAddEndPuncttrue
\mciteSetBstMidEndSepPunct{\mcitedefaultmidpunct}
{\mcitedefaultendpunct}{\mcitedefaultseppunct}\relax
\EndOfBibitem
\bibitem[Transform()]{transform_GaN}
Transform TP90H050WS Datasheet. \url{https://www.transphormusa.com/en/document/datasheet-tp90h050ws/}\relax
\mciteBstWouldAddEndPuncttrue
\mciteSetBstMidEndSepPunct{\mcitedefaultmidpunct}
{\mcitedefaultendpunct}{\mcitedefaultseppunct}\relax
\EndOfBibitem
\bibitem[Meneghini \latin{et~al.}(2021)Meneghini, De~Santi, Abid, Buffolo, Cioni, Khadar, Nela, Zagni, Chini, Medjdoub, Meneghesso, Verzellesi, Zanoni, and Matioli]{GaN_status_2021}
Meneghini,~M.; De~Santi,~C.; Abid,~I.; Buffolo,~M.; Cioni,~M.; Khadar,~R.~A.; Nela,~L.; Zagni,~N.; Chini,~A.; Medjdoub,~F.; Meneghesso,~G.; Verzellesi,~G.; Zanoni,~E.; Matioli,~E. {GaN-based power devices: Physics, reliability, and perspectives}. \emph{Journal of Applied Physics} \textbf{2021}, \emph{130}, 181101\relax
\mciteBstWouldAddEndPuncttrue
\mciteSetBstMidEndSepPunct{\mcitedefaultmidpunct}
{\mcitedefaultendpunct}{\mcitedefaultseppunct}\relax
\EndOfBibitem
\bibitem[Bhattacharyya \latin{et~al.}(2022)Bhattacharyya, Sharma, Alema, Ranga, Roy, Peterson, Seryogin, Osinsky, Singisetti, and Krishnamoorthy]{Bhattacharyya_2022}
Bhattacharyya,~A.; Sharma,~S.; Alema,~F.; Ranga,~P.; Roy,~S.; Peterson,~C.; Seryogin,~G.; Osinsky,~A.; Singisetti,~U.; Krishnamoorthy,~S. 4.4 kV $\beta$-Ga$_2$O$_3$ MESFETs with power figure of merit exceeding 100 MW.cm$^2$. \emph{Applied Physics Express} \textbf{2022}, \emph{15}, 061001\relax
\mciteBstWouldAddEndPuncttrue
\mciteSetBstMidEndSepPunct{\mcitedefaultmidpunct}
{\mcitedefaultendpunct}{\mcitedefaultseppunct}\relax
\EndOfBibitem
\bibitem[Biswas \latin{et~al.}(2019)Biswas, Joishi, Biswas, Thakar, Rajan, and Lodha]{Biswas2019}
Biswas,~D.; Joishi,~C.; Biswas,~J.; Thakar,~K.; Rajan,~S.; Lodha,~S. {Enhanced n-type $\beta$-Ga$_2$O$_3$ ($\bar{2}$01) gate stack performance using Al$_2$O$_3$/SiO$_2$ bi-layer dielectric}. \emph{Applied Physics Letters} \textbf{2019}, \emph{114}, 212106\relax
\mciteBstWouldAddEndPuncttrue
\mciteSetBstMidEndSepPunct{\mcitedefaultmidpunct}
{\mcitedefaultendpunct}{\mcitedefaultseppunct}\relax
\EndOfBibitem
\bibitem[Sharma and Lodha(2024)Sharma, and Lodha]{psharma2024}
Sharma,~P.; Lodha,~S. { $\beta$-Ga$_2$O$_3$ Schottky barrier height improvement using Ar/O$_2$ plasma and HF surface treatments}. \emph{Applied Physics Letters} \textbf{2024}, \emph{124}, 072106\relax
\mciteBstWouldAddEndPuncttrue
\mciteSetBstMidEndSepPunct{\mcitedefaultmidpunct}
{\mcitedefaultendpunct}{\mcitedefaultseppunct}\relax
\EndOfBibitem
\bibitem[Sharma \latin{et~al.}(2024)Sharma, Parasubotu, and Lodha]{PSedtm2024}
Sharma,~P.; Parasubotu,~Y.; Lodha,~S. High-k dielectric integration to improve breakdown characteristics of $\beta$-Ga$_2$O$_3$ Schottky diode. 2024 8th IEEE Electron Devices Technology I\& Manufacturing Conference (EDTM). 2024; pp 1--3\relax
\mciteBstWouldAddEndPuncttrue
\mciteSetBstMidEndSepPunct{\mcitedefaultmidpunct}
{\mcitedefaultendpunct}{\mcitedefaultseppunct}\relax
\EndOfBibitem
\bibitem[Higashiwaki \latin{et~al.}(2012)Higashiwaki, Sasaki, Kuramata, Masui, and Yamakoshi]{Masataka_2012_RIE_etch}
Higashiwaki,~M.; Sasaki,~K.; Kuramata,~A.; Masui,~T.; Yamakoshi,~S. {Gallium oxide (Ga$_2$O$_3$) metal-semiconductor field-effect transistors on single-crystal β-Ga$_2$O$_3$ (010) substrates}. \emph{Applied Physics Letters} \textbf{2012}, \emph{100}, 013504\relax
\mciteBstWouldAddEndPuncttrue
\mciteSetBstMidEndSepPunct{\mcitedefaultmidpunct}
{\mcitedefaultendpunct}{\mcitedefaultseppunct}\relax
\EndOfBibitem
\bibitem[Biswas \latin{et~al.}(2020)Biswas, Joishi, Biswas, Tiwari, and Lodha]{biswas2020}
Biswas,~D.; Joishi,~C.; Biswas,~J.; Tiwari,~P.; Lodha,~S. {Charge trap layer enabled positive tunable V$_{FB}$ in $\beta$-Ga$_2$O$_3$ gate stacks for enhancement mode transistors}. \emph{Applied Physics Letters} \textbf{2020}, \emph{117}, 172101\relax
\mciteBstWouldAddEndPuncttrue
\mciteSetBstMidEndSepPunct{\mcitedefaultmidpunct}
{\mcitedefaultendpunct}{\mcitedefaultseppunct}\relax
\EndOfBibitem
\bibitem[Singh \latin{et~al.}(2018)Singh, Casbon, Uren, Pomeroy, Dalcanale, Karboyan, Tasker, Wong, Sasaki, Kuramata, Yamakoshi, Higashiwaki, and Kuball]{MKuball2018_Vth}
Singh,~M.; Casbon,~M.~A.; Uren,~M.~J.; Pomeroy,~J.~W.; Dalcanale,~S.; Karboyan,~S.; Tasker,~P.~J.; Wong,~M.~H.; Sasaki,~K.; Kuramata,~A.; Yamakoshi,~S.; Higashiwaki,~M.; Kuball,~M. Pulsed Large Signal RF Performance of Field-Plated Ga$_2$O$_3$ MOSFETs. \emph{IEEE Electron Device Letters} \textbf{2018}, \emph{39}, 1572--1575\relax
\mciteBstWouldAddEndPuncttrue
\mciteSetBstMidEndSepPunct{\mcitedefaultmidpunct}
{\mcitedefaultendpunct}{\mcitedefaultseppunct}\relax
\EndOfBibitem
\bibitem[Wong \latin{et~al.}(2016)Wong, Sasaki, Kuramata, Yamakoshi, and Higashiwaki]{MHWong2016_Vth}
Wong,~M.~H.; Sasaki,~K.; Kuramata,~A.; Yamakoshi,~S.; Higashiwaki,~M. Field-Plated Ga$_2$O$_3$ MOSFETs With a Breakdown Voltage of Over 750 V. \emph{IEEE Electron Device Letters} \textbf{2016}, \emph{37}, 212--215\relax
\mciteBstWouldAddEndPuncttrue
\mciteSetBstMidEndSepPunct{\mcitedefaultmidpunct}
{\mcitedefaultendpunct}{\mcitedefaultseppunct}\relax
\EndOfBibitem
\bibitem[Higashiwaki \latin{et~al.}(2013)Higashiwaki, Sasaki, Wong, Kamimura, Krishnamurthy, Kuramata, Masui, and Yamakoshi]{MHWong2013_Vth_BV}
Higashiwaki,~M.; Sasaki,~K.; Wong,~M.~H.; Kamimura,~T.; Krishnamurthy,~D.; Kuramata,~A.; Masui,~T.; Yamakoshi,~S. Depletion-mode Ga$_2$O$_3$ MOSFETs on $\beta$-Ga$_2$O$_3$ (010) substrates with Si-ion-implanted channel and contacts. 2013 IEEE International Electron Devices Meeting. 2013; pp 28.7.1--28.7.4\relax
\mciteBstWouldAddEndPuncttrue
\mciteSetBstMidEndSepPunct{\mcitedefaultmidpunct}
{\mcitedefaultendpunct}{\mcitedefaultseppunct}\relax
\EndOfBibitem
\bibitem[Konishi \latin{et~al.}(2016)Konishi, Kamimura, Wong, Sasaki, Kuramata, Yamakoshi, and Higashiwaki]{konishi2016large}
Konishi,~K.; Kamimura,~T.; Wong,~M.~H.; Sasaki,~K.; Kuramata,~A.; Yamakoshi,~S.; Higashiwaki,~M. Large conduction band offset at SiO$_2$/$\beta$-Ga$_2$O$_3$ heterojunction determined by x-ray photoelectron spectroscopy. \emph{physica status solidi (b)} \textbf{2016}, \emph{253}, 623--625\relax
\mciteBstWouldAddEndPuncttrue
\mciteSetBstMidEndSepPunct{\mcitedefaultmidpunct}
{\mcitedefaultendpunct}{\mcitedefaultseppunct}\relax
\EndOfBibitem
\bibitem[Zeng \latin{et~al.}(2016)Zeng, Jia, and Singisetti]{zeng2016interface}
Zeng,~K.; Jia,~Y.; Singisetti,~U. Interface State Density in Atomic Layer Deposited SiO$_2$/$\beta$-Ga$_2$O$_3$ ($\bar{2}$01) MOSCAPs. \emph{IEEE Electron Device Letters} \textbf{2016}, \emph{37}, 906--909\relax
\mciteBstWouldAddEndPuncttrue
\mciteSetBstMidEndSepPunct{\mcitedefaultmidpunct}
{\mcitedefaultendpunct}{\mcitedefaultseppunct}\relax
\EndOfBibitem
\bibitem[Zeng \latin{et~al.}(2018)Zeng, Vaidya, and Singisetti]{Uttam2018_BV}
Zeng,~K.; Vaidya,~A.; Singisetti,~U. 1.85 kV breakdown voltage in lateral field-plated Ga$_2$O$_3$ MOSFETs. \emph{IEEE Electron Device Letters} \textbf{2018}, \emph{39}, 1385--1388\relax
\mciteBstWouldAddEndPuncttrue
\mciteSetBstMidEndSepPunct{\mcitedefaultmidpunct}
{\mcitedefaultendpunct}{\mcitedefaultseppunct}\relax
\EndOfBibitem
\end{mcitethebibliography}


\providecommand{\noopsort}[1]{}\providecommand{\singleletter}[1]{#1}%
\providecommand{\latin}[1]{#1}
\makeatletter
\providecommand{\doi}
  {\begingroup\let\do\@makeother\dospecials
  \catcode`\{=1 \catcode`\}=2 \doi@aux}
\providecommand{\doi@aux}[1]{\endgroup\texttt{#1}}
\makeatother
\providecommand*\mcitethebibliography{\thebibliography}
\csname @ifundefined\endcsname{endmcitethebibliography}  {\let\endmcitethebibliography\endthebibliography}{}
\begin{mcitethebibliography}{7}
\providecommand*\natexlab[1]{#1}
\providecommand*\mciteSetBstSublistMode[1]{}
\providecommand*\mciteSetBstMaxWidthForm[2]{}
\providecommand*\mciteBstWouldAddEndPuncttrue
  {\def\EndOfBibitem{\unskip.}}
\providecommand*\mciteBstWouldAddEndPunctfalse
  {\let\EndOfBibitem\relax}
\providecommand*\mciteSetBstMidEndSepPunct[3]{}
\providecommand*\mciteSetBstSublistLabelBeginEnd[3]{}
\providecommand*\EndOfBibitem{}
\mciteSetBstSublistMode{f}
\mciteSetBstMaxWidthForm{subitem}{(\alph{mcitesubitemcount})}
\mciteSetBstSublistLabelBeginEnd
  {\mcitemaxwidthsubitemform\space}
  {\relax}
  {\relax}

\bibitem[Koblmüller \latin{et~al.}(2010)Koblmüller, Chu, Raman, Mishra, and Speck]{kobli2010_AlGaN/GaN}
Koblmüller,~G.; Chu,~R.~M.; Raman,~A.; Mishra,~U.~K.; Speck,~J.~S. {High-temperature molecular beam epitaxial growth of AlGaN/GaN on GaN templates with reduced interface impurity levels}. \emph{Journal of Applied Physics} \textbf{2010}, \emph{107}, 043527\relax
\mciteBstWouldAddEndPuncttrue
\mciteSetBstMidEndSepPunct{\mcitedefaultmidpunct}
{\mcitedefaultendpunct}{\mcitedefaultseppunct}\relax
\EndOfBibitem
\bibitem[Izumi \latin{et~al.}(1993)Izumi, Yoshida, Takano, Nishitani, and Otsubo]{IZUMI1993123}
Izumi,~S.; Yoshida,~N.; Takano,~H.; Nishitani,~K.; Otsubo,~M. Study on the accumulated impurities at the epilayer/substrate interface and their influence on the leakage current of metal-semiconductor-field effect transistors. \emph{Journal of Crystal Growth} \textbf{1993}, \emph{133}, 123--131\relax
\mciteBstWouldAddEndPuncttrue
\mciteSetBstMidEndSepPunct{\mcitedefaultmidpunct}
{\mcitedefaultendpunct}{\mcitedefaultseppunct}\relax
\EndOfBibitem
\bibitem[Bhattacharyya \latin{et~al.}(2023)Bhattacharyya, Peterson, Itoh, Roy, Cooke, Rebollo, Ranga, Sensale-Rodriguez, and Krishnamoorthy]{AB_2023_Sidoping_nonidealities}
Bhattacharyya,~A.; Peterson,~C.; Itoh,~T.; Roy,~S.; Cooke,~J.; Rebollo,~S.; Ranga,~P.; Sensale-Rodriguez,~B.; Krishnamoorthy,~S. {Enhancing the electron mobility in Si-doped (010) $\beta$-Ga$_2$O$_3$ films with low-temperature buffer layers}. \emph{APL Materials} \textbf{2023}, \emph{11}, 021110\relax
\mciteBstWouldAddEndPuncttrue
\mciteSetBstMidEndSepPunct{\mcitedefaultmidpunct}
{\mcitedefaultendpunct}{\mcitedefaultseppunct}\relax
\EndOfBibitem
\bibitem[Kumar \latin{et~al.}(2020)Kumar, Kamimura, Lin, Nakata, and Higashiwaki]{Masataka_leakagechannel_2020}
Kumar,~S.; Kamimura,~T.; Lin,~C.-H.; Nakata,~Y.; Higashiwaki,~M. {Reduction in leakage current through interface between Ga$_2$O$_3$ epitaxial layer and substrate by ion implantation doping of compensating impurities}. \emph{Applied Physics Letters} \textbf{2020}, \emph{117}, 193502\relax
\mciteBstWouldAddEndPuncttrue
\mciteSetBstMidEndSepPunct{\mcitedefaultmidpunct}
{\mcitedefaultendpunct}{\mcitedefaultseppunct}\relax
\EndOfBibitem
\bibitem[Tetzner \latin{et~al.}(2019)Tetzner, Bahat~Treidel, Hilt, Popp, Bin~Anooz, Wagner, Thies, Ickert, Gargouri, and Würfl]{tetzner2019lateral}
Tetzner,~K.; Bahat~Treidel,~E.; Hilt,~O.; Popp,~A.; Bin~Anooz,~S.; Wagner,~G.; Thies,~A.; Ickert,~K.; Gargouri,~H.; Würfl,~J. Lateral 1.8 kV $\beta$-Ga$_2$O$_3$ MOSFET With 155 MW/cm$^2$ Power Figure of Merit. \emph{IEEE Electron Device Letters} \textbf{2019}, \emph{40}, 1503--1506\relax
\mciteBstWouldAddEndPuncttrue
\mciteSetBstMidEndSepPunct{\mcitedefaultmidpunct}
{\mcitedefaultendpunct}{\mcitedefaultseppunct}\relax
\EndOfBibitem
\bibitem[Tetzner \latin{et~al.}(2020)Tetzner, Hilt, Popp, Anooz, and Würfl]{tetzner2020challenges}
Tetzner,~K.; Hilt,~O.; Popp,~A.; Anooz,~S.~B.; Würfl,~J. Challenges to overcome breakdown limitations in lateral $\beta$-Ga$_2$O$_3$ MOSFET devices. \emph{Microelectronics Reliability} \textbf{2020}, \emph{114}, 113951\relax
\mciteBstWouldAddEndPuncttrue
\mciteSetBstMidEndSepPunct{\mcitedefaultmidpunct}
{\mcitedefaultendpunct}{\mcitedefaultseppunct}\relax
\EndOfBibitem
\end{mcitethebibliography}

\end{document}


\maketitle
\newpage
\section{MOSFET hysteresis}
\FloatBarrier
\begin{figure*}[h]

\includegraphics[width=0.5\linewidth]{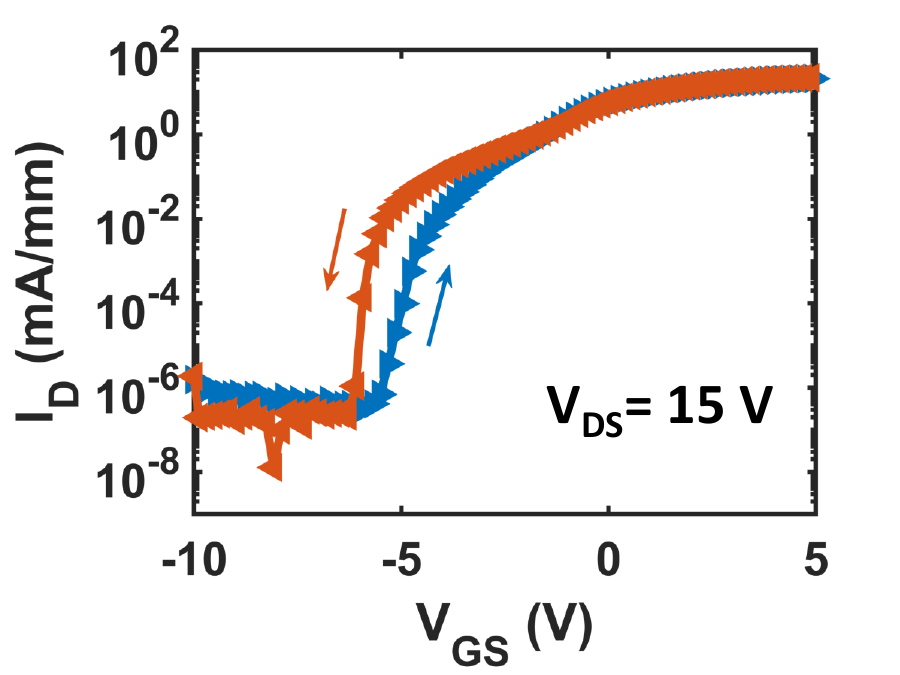}
\caption{\label{fig:fig1}Single-gate MOSFET transfer characteristics showing hysteresis in double-sweep (forward and reverse) $I-V$ measurements. }
\end{figure*}
\FloatBarrier

Non-ideal transfer characteristics with a double hump/kink as shown in Fig. \ref{fig:fig1}  are likely caused by a parasitic conduction path, activated at a different threshold voltage than the surface channel, at the $\beta$-Ga$_2$O$_3$ substrate-epi-layer interface\cite{kobli2010_AlGaN/GaN, IZUMI1993123}. Residual Si doping at the interface of Si-doped epi-layers and un-doped $\beta$-Ga$_2$O$_3$ substrates\cite{AB_2023_Sidoping_nonidealities, Masataka_leakagechannel_2020} increase leakage in lateral transistors, negatively impacting device performance resulting in premature breakdown, compromising reliability, and shortening longevity  \cite{tetzner2019lateral,tetzner2020challenges}.

\section{Reverse breakdown characteristics of single-gate unidirectional MOSFET}

\begin{figure*}[h]

\includegraphics[width=0.5\linewidth]{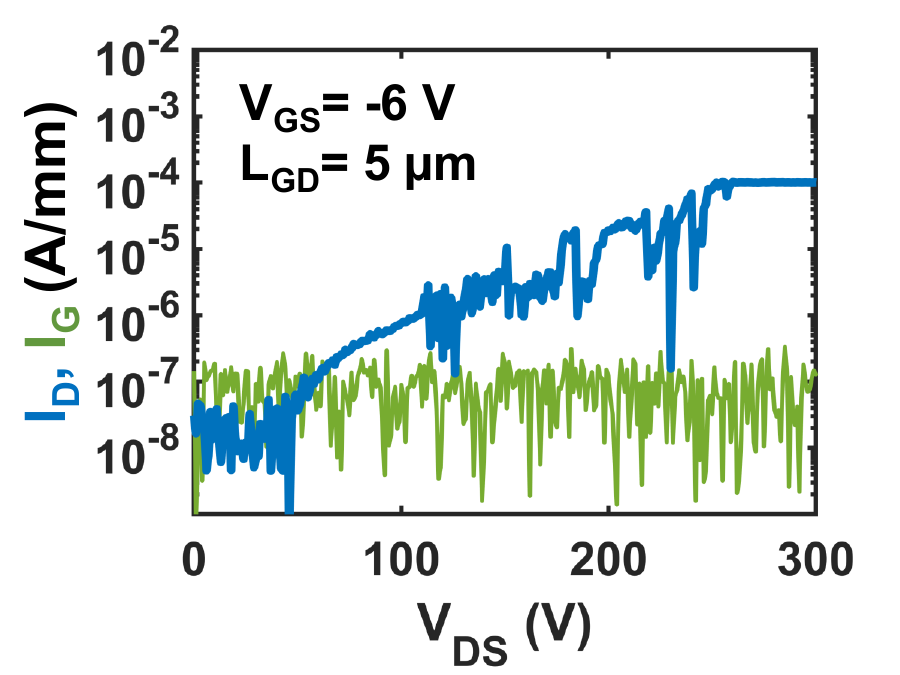}
\caption{\label{fig:fig2}Reverse bias $I-V$ characteristics of single-gate unidirectional MOSFET showing $V_{BR}$ > 250 V with current compliance of 0.1 mA/mm. }
\end{figure*}

The reverse $I-V$ characteristics of single-gate unidirectional MOSFET were measured by sweeping drain-to-source voltage ($V_{DS}$) from 0 to -300 V keeping the current compliance at 0.1 mA/mm and gate bias ($V_{GS}$) at -6 V. The $I-V$ characteristics shown in Fig. \ref{fig:fig2} indicate a breakdown voltage ($V_{BR}$) value of more than 250 V.

\FloatBarrier

\section{Transfer length measurement (TLM): contact and sheet resistance}
\FloatBarrier
\begin{figure*}[h]

\includegraphics[width=1\linewidth]{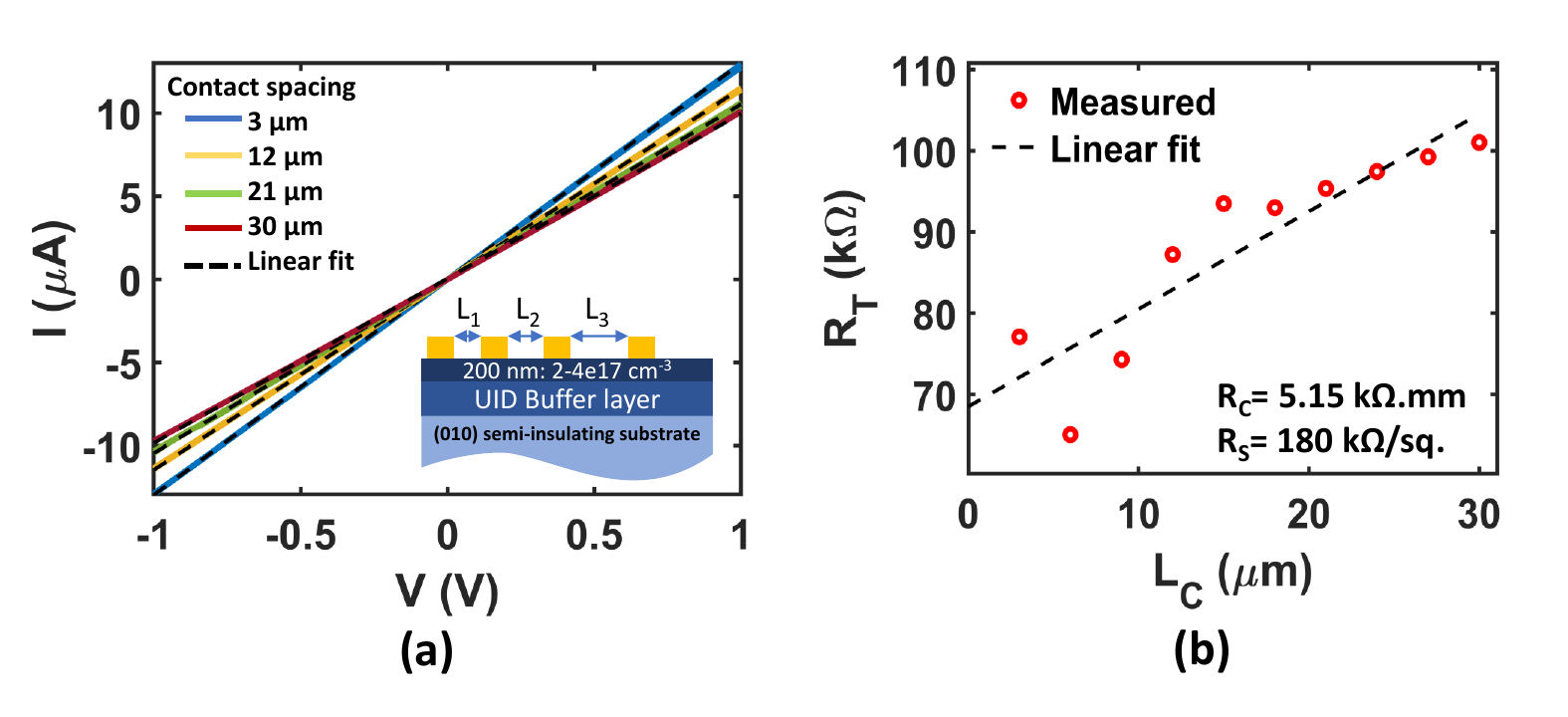}
\caption{\label{fig:fig3} (a) $I-V$ data measured for different sets of contacts in the TLM architecture, along with the device cross-section, (b) total resistance measured vs. contact spacing used for extracting contact and sheet resistance based on the TLM measurement technique.}
\end{figure*}
\FloatBarrier
For 200 nm thick and $1-5$ x $10^{17}$ cm$^{-3}$ doped epi layer, the extracted sheet resistance and contact resistance values are 180 k$\Omega$/sq. and 5.15 k$\Omega$.mm respectively from the TLM data shown in Fig. \ref{fig:fig3}. 

\bibliography{manuscript}